\documentclass[aps,prd,reprint,superscriptaddress,amsmath,amssymb,nofootinbib,floatfix]{revtex4-2}
\usepackage{graphicx}
\usepackage{dcolumn}
\usepackage[utf8]{inputenc}

\usepackage{ulem}
\usepackage{xcolor}
\usepackage[unicode=true,colorlinks,linkcolor=blue,anchorcolor=blue,urlcolor=blue,citecolor=blue,breaklinks=true]{hyperref}
\usepackage{orcidlink}
\usepackage{subfigure}

\newcommand{\pararrow}{\mathord{\buildrel{\lower3pt\hbox{$\scriptscriptstyle\leftrightarrow$}}\over {\partial}}} 
\newcommand{\pararrowk}[1]{\mathord{\buildrel{\lower3pt\hbox{$\scriptscriptstyle\leftrightarrow$}}\over {\partial}\hspace*{-0.18em}{}^#1}\hspace*{-0.18em} \,} 

\usepackage{physics}
\usepackage{bm}

\newcommand{\qfnu}{\affiliation{College of Physics and Engineering, Qufu Normal University, Qufu 273165, China}}

\newcommand{\itp}{\affiliation{Institute of Theoretical Physics, Chinese Academy of Sciences, Beijing 100190, China}}

\newcommand{\scnt}{\affiliation{Southern Center for Nuclear-Science Theory (SCNT), Institute of Modern Physics, Chinese Academy of Sciences, Huizhou 516000, China}}

\begin{document}
\title{ Radiative decays of $X(3872)$  within $D{\bar D}^*$ molecular framework  }

\author{Hao-Dong Cai\,\orcidlink{0009-0000-5269-8621}} \qfnu
\author{Run-Hao Chen\,\orcidlink{0009-0005-4718-4825}} \qfnu
\author{Yuan-Jun Gao\,} \qfnu
\author{Zhao-Sai Jia\,\orcidlink{0000-0002-7133-189X}}~\email{jzsqfphys@163.com} \scnt \itp 
\author{Gang Li\,\orcidlink{0000-0002-5227-8296}}~\email{gli@qfnu.edu.cn} \qfnu
\author{Shi-Dong Liu\,\orcidlink{0000-0001-9404-5418}}~\email{liusd@qfnu.edu.cn} \qfnu

\begin{abstract}
Within the framework of nonrelativistic effective field theory, we calculate the radiative decay widths of the process $X(3872) \to D\bar{D} \gamma$ while taking into account the $D\bar{D}$ final-state interactions via resonance saturation. The $X(3872)$ is treated as a $D^*\bar{D} +\rm c.c.$ bound state with $J^{PC}=1^{++}$. For a representative mixing angle $\theta=\pi/4$, our numerical calculations predict that the tree-level partial decay width for $X(3872) \to D^0 \bar{D}^0\gamma$ can reach approximately $7.3$ keV, while that for $X(3872) \to D^+ D^- \gamma$ is less than $1$ keV. It is found that the contribution of the triangle loop diagrams to the total decay width of $X(3872) \to  D^0 \bar{D}^0\gamma$ is very small, amounting to less than $1\%$ of the tree-level result. For the $X(3872)\to  D^+  D^-\gamma $ decay, the contribution from the triangle loop diagrams is relatively more significant, enhancing the total decay width by roughly $1.3\%$ relative to the tree-level result. We expect that the present predictions based on the molecular picture of the $X(3872)$ can be tested by future experiments.
\end{abstract}

\date{\today}
\maketitle

\section{Introduction}
\label{sec:introduction}

Since 2003, a large number of exotic hadron states have been observed in high-energy experiments and have also been predicted in theoretical investigations. Their mass spectra, quantum numbers and decay widths differ substantially from the predictions of the conventional quark model. Such exotic candidates are generally referred to as $XYZ$ states, which have attracted extensive experimental and theoretical efforts to explore their internal structure, for reviews see Refs.~\cite{Richard:2016eis,Hosaka:2016pey,Guo:2017jvc,Ali:2017jda,Liu:2019zoy,Chen:2022asf}. Among the observed $XYZ$ states, $X(3872)$ is the earliest discovered and remains the most comprehensively studied one. It was first reported by the Belle Collaboration in 2003 through the $J/\psi \pi^+ \pi^-$ invariant mass spectrum~\cite{Belle:2003nnu}. In 2013, its spin-parity quantum numbers were determined to be $J^{PC}=1^{++}$ through the decay process $B^+\to X(3872) K^+ \to \pi^+\pi^- J/\psi K^+\to \pi^+\pi^-\mu^+\mu^- K^+$~\cite{LHCb:2013kgk}. The current world-averaged mass and total decay width of $X(3872)$ are $m_{X}=(3871.64\pm 0.06)$ MeV and $\Gamma=(1.19 \pm 0.21)$ MeV, respectively~\cite{ParticleDataGroup:2024cfk}. Given that its mass lies extremely close to the $D^0\bar{D}^{*0}$ threshold, $X(3872)$ is widely interpreted as a promising $D\bar{D}^{*}$ hadronic molecular state~\cite{Close:2003sg,Pakvasa:2003ea,Swanson:2003tb,Swanson:2004pp,Tornqvist:2004qy,Voloshin:2003nt,Wong:2003xk,AlFiky:2005jd,Braaten:2006sy,Fleming:2007rp,Ding:2009vj,Dong:2009yp,Lee:2009hy,Lee:2011rka,Liu:2009qhy,Zhang:2009vs,Gamermann:2009uq,Mehen:2011ds,Nieves:2011zz,Nieves:2012tt,Li:2012cs,Sun:2012sy,Guo:2013sya,He:2014nya,Zhao:2014gqa,Guo:2014taa,Guo:2014hqa,Braaten:2003he,Wu:2021udi,Yamaguchi:2019vea,Wang:2025zss,Cai:2025inq,Wu:2025crk,Jia:2025xil}. Comprehensive reviews focusing on the molecular interpretation of $X(3872)$ can be found in Refs.~\cite{Guo:2017jvc,Kalashnikova:2018vkv}. Nevertheless, alternative structural interpretations have also been proposed, including the tetraquark configuration~\cite{Maiani:2004vq,Maiani:2005pe,Maiani:2007vr,Terasaki:2007uv,Dubnicka:2020yxy,Wang:2023sii,Wang:2019tlw} and the conventional charmonium assignment~\cite{Barnes:2003vb,Suzuki:2005ha}.

Investigating the diverse decay modes of $X(3872)$ is essential for clarifying its intrinsic properties and internal structure, see Refs.~\cite{Fleming:2007rp, DeFazio:2008xq,Dong:2009uf,Guo:2014taa,Dai:2019hrf,Colangelo:2025uhs,Wang:2025zss,Nakamura:2026xax}. Within the molecular framework and effective field theory, Ref.~\cite{Guo:2014taa} analyzed the radiative decays $X(3872)\to \gamma J/\psi$ and $X(3872)\to \gamma \psi'$. The results demonstrated that the experimental ratio $R=\mathcal{B}(X(3872)\to\gamma \psi')/\mathcal{B}(X(3872)\to\gamma J/\psi)$ is consistent with the wave function of $X(3872)$ dominated by the $D\bar{D}^*$ hadronic molecular component. To discriminate between a pure molecular state scenario and a mixed scenario with substantial $c\bar{c}$ charmonium admixture, it is crucial to clarify the interplay between charmonium and molecular components. The analysis in Ref.~\cite{Dong:2009uf} showed that the ratio $\Gamma(X(3872)\to\gamma J/\psi)/\Gamma(X(3872)\to \pi^+\pi^-J/\psi)$, which relates radiative and strong decays, indicates that the $c\bar{c}$ charmonium component plays only a subleading role in the structure of $X(3872)$. Recently, under the assumption that $X(3872)$ is a pure hadronic molecular state composed of $D\bar{D}^* +\rm c.c.$, Ref.~\cite{Wang:2025zss} investigated the radiative decays $X(3872)\to \gamma \rho^0$ and $X(3872) \to \gamma \omega$. These decay modes are highly sensitive to the molecular configuration governed by the relative proportion of neutral and charged components. 

XEFT, as a nonrelativistic effective field theory, was originally proposed to investigate the long-range properties of the $X(3872)$ state. Its effective degrees of freedom include $D^0$, $D^{*0}$, $\bar{D}^0$, $\bar{D}^{*0}$, and $\pi^0$, all treated nonrelativistically~\cite{Fleming:2007rp}. More applications of XEFT are discussed in Refs.~\cite{Braaten:2010mg,Fleming:2011xa,Mehen:2015efa,Braaten:2015tga,Dai:2019hrf}. In particular, Ref.~\cite{Dai:2019hrf} revisited the decay process $X(3872)\to D^0\bar{D}^0 \pi^0$ within the XEFT framework, incorporating final-state interaction (FSI) diagrams for $\pi^0 D^0$, $\pi^0 \bar{D}^0$, and $D^0\bar{D}^0$ systems. The analysis revealed that $D^0\bar{D}^0$ FSI dominates the theoretical uncertainty, contributing approximately half of the total uncertainty, with the overall uncertainty of the partial width reaching around $50\%$.

Experimentally, the BESIII Collaboration set upper limits at $90\%$ confidence level on the branching fraction ratios relative to the decay mode $X(3872)\to \pi^+\pi^-J/\psi$, $\mathcal{B}(X(3872) \to \bar{D}^0 D^0\gamma)/\mathcal{B}(X(3872)\to \pi^+\pi^-J/\psi)<1.58$ and $\mathcal{B}(X(3872) \to D^- D^+\gamma)/\mathcal{B}(X(3872)\to \pi^+\pi^-J/\psi)<0.99$~\cite{BESIII:2020nbj}. However, no significant signal was observed for these radiative decay channels. In Refs.~\cite{Voloshin:2005rt, Colangelo:2007ph}, the radiative decay $X(3872) \to D\bar{D}\gamma$ was proposed as a probe of the $X(3872)$ internal structure, predicting a small ratio $R = \Gamma(X \to D^+D^-\gamma)/\Gamma(X \to D^0\bar{D}^0\gamma) \ll 1$ for a molecular interpretation. However, these works were based on tree-level amplitudes and neglected the $D\bar{D}$ FSI effects, which have been systematically discussed in a recent review~\cite{Wang:2025dur}. To better understand the nature of $X(3872)$, we employ XEFT to systematically study its radiative decays $X(3872)\to D \bar{D}\gamma$, incorporating the $D\bar{D}$ FSI effects. In our calculations, the $X(3872)$ is assumed to be a molecular state with $J^{PC}=1^{++}$, composed of $D^{*0}\bar{D}^0 + D^{*+} D^-$ and their charge conjugates.

This paper is structured as follows. In Sec~\ref{sec:formalism}, we introduce the effective Lagrangians of the relevant vertices, and provide the corresponding decay amplitudes. Numerical results and discussions are given in Sec~\ref{sec:results}. We conclude with a brief summary in the last section, while detailed formulas are provided in the Appendix.

\section{Theoretical Framework}\label{sec:formalism}
\subsection{Effective Lagrangians}
\label{subsec:effective-L}

\begin{figure*}
	\centering
	\includegraphics[width=0.85\linewidth]{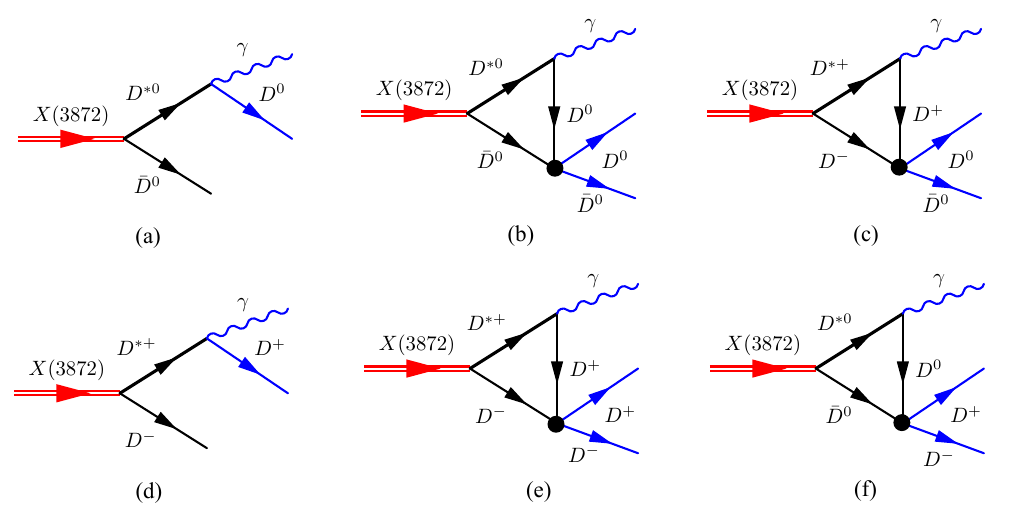}\\
    \caption{Feynman diagrams for the decay $X(3872) \to  D \bar{D}\gamma$.  The charge conjugated loops are not shown here, but included in our calculations.}
    \label{fig:feyndiags}
\end{figure*}

In this section, we present the relevant Feynman diagrams and effective Lagrangians for the decay $X(3872) \to D \bar{D}\gamma$. The Feynman diagrams are shown in Fig.~\ref{fig:feyndiags}, where Figures~\ref{fig:feyndiags}(a)--\ref{fig:feyndiags}(c) depict the process $X(3872) \to  D^0 \bar{D}^0\gamma$, and figures~\ref{fig:feyndiags}(d)--\ref{fig:feyndiags}(f) correspond to the decay $X(3872) \to  D^+ D^-\gamma$.

We assume that $X(3872)$ is an S-wave molecular state with $J^{PC}=1^{++}$, containing a superposition of the $D^{*0}\bar{D}^0$ and $D^{*+}D^-$ two-hadron configurations~\cite{Wang:2022qxe,Guo:2013zbw}
\begin{eqnarray}\label{eq:x3872wf}
\left|X(3872)\right\rangle &=& \frac{\cos{\theta}}{\sqrt{2}}(D^{\ast 0} \bar{D}^0+D^0 \bar{D}^{\ast0})\nonumber\\ &+& \frac{\sin{\theta}}{\sqrt{2}}(D^{\ast + } {D}^{-}+D^+ D^{\ast -})\,.
\end{eqnarray}
Here, $\theta$ denotes a phase angle describing the relative proportion of the neutral and charged constituent components. The effective Lagrangian describing the coupling between $X(3872)$ and $D^{*}\bar{D}$ can be written as ~\cite{Wang:2022qxe} 
\begin{align}\label{eq:X3872Lagrangian} 
 \mathcal{L}_{X} &= \frac{g_n}{\sqrt{2}}X^{\dagger}_{i}(D^{\ast 0 i} \bar{D}^0+D^0 \bar{D}^{\ast0 i})\nonumber\\ &+ \frac{g_c}{\sqrt{2}}X^{\dagger}_{i}(D^{\ast + i} {D}^{-}+D^+ D^{\ast - i })\,,
\end{align}
where $g_n$ and $g_c$ are the coupling constants of the $X(3872)$ to neutral and charged charmed mesons, respectively. These coupling constants can be extracted from the residues of the $D^{0}\bar{D}^{*0} - D^{+}D^{*-}$ coupled-channel scattering $T$ matrix at the $X(3872)$ pole position as~\cite{Albaladejo:2015dsa,Jia:2025xil}
\begin{eqnarray}\label{eq:XDDsct}
    g_n=\sqrt{\frac{2\pi \gamma_n}{\mu ^2_n}} \cos{\theta},\quad
    g_c=\sqrt{\frac{2\pi \gamma_c}{\mu^2_c}} \sin{\theta},
\end{eqnarray}
where $\gamma_n =\sqrt{2\mu_n E_n}$ and $\gamma_c =\sqrt{2\mu_c E_c}$. $E_n=m_{D^{*0}}+m_{\bar{D}^0}-m_{X}$ and $E_c=m_{D^{*+}}+m_{D^-}-m_{X}=\Delta+E_n$ are the binding energies of the $X(3872)$ relative to the neutral and charged thresholds, respectively, $\Delta=m_{D^{*+}}+m_{D^-}-m_{D^{*0}}-m_{\bar{D}^0}$, $\mu_n$ and $\mu_c$ are the reduced masses of $D^{*0}\bar{D}^{0}$ and $D^{*+}D^-$, respectively.

To systematically describe the decay mechanism of $X(3872) \to D \bar{D}\gamma$, we employ the general effective Lagrangians for the interactions between charmed mesons and photons, as well as the contact interactions among charmed mesons, which are given by~\cite{Dai:2019hrf,Jia:2023hvc,Guo:2017jvc,Fleming:2021wmk,Amundson:1992yp,Stewart:1998ke}
\begin{eqnarray}\label{eq:XEFTeL}
    \mathcal{L}&=&D_i^{* \dagger} \left( i\partial^0 + \frac{\nabla^2}{2m_{D^*}} -\delta \right) D^{*i}
+ D^{\dagger} \left( i\partial^0 + \frac{\nabla^2}{2m_{D}} \right) D\nonumber \\
&+&\bar{D}_i^{* \dagger} \left( i\partial^0 + \frac{\nabla^2}{2m_{D^*}} -\delta \right) \bar{D}^{*i}
+ \bar{D}^{\dagger} \left( i\partial^0 + \frac{\nabla^2}{2m_{D}} \right) \bar{D}\nonumber \\
&-& \frac{C_0}{\sqrt{2}} (\bar{D}^*D + D^*\bar{D})^\dagger \cdot (\bar{D}^*D + D^*\bar{D})\nonumber\\
&+&\frac{1}{2}(D^\dagger \mu_D \mathcal{B}^i D^{*i}+\text{H.c.})+\frac{1}{2}C_{0D}D^\dagger \bar{D}^\dagger D \bar{D}\,,
\end{eqnarray}
where $D^{(*)}=(D^{(*)0},D^{(*)+})$ denotes the pseudoscalar (vector) charmed mesons and $\bar{D}^{(*)}$ represents the corresponding anticharmed mesons. $\delta=m_{D^*}-m_{D}$, and the magnetic field $\mathcal{B}^k=\epsilon^{ijk} \partial^i A^j$. In this Lagrangian, the first two lines contain the kinetic terms for the charmed mesons. The third line contains the contact interactions between $D(\bar{D})$ and $\bar{D}^*(D^*)$ in the $I=0$ channel. In the last line, we present the magnetic couplings between photons and charmed mesons, and the contact interactions for $D\bar{D}\to D\bar{D}$. Here $\mu_D=\rm diag(\mu_{D^0},\mu_{D^+})$ is the matrix of the transition magnetic moments, where $\mu_{D^0}=0.56\, \rm GeV^{-1}$ and $\mu_{D^+}=-0.15\, \rm GeV^{-1}$ are extracted from the decay widths $\Gamma[D^{*0}\to D^0 \gamma ]=19.52$ keV~\cite{Guo:2019qcn} and $\Gamma[D^{*+}\to D^+ \gamma ]=1.33$ keV~\cite{ParticleDataGroup:2024cfk}.

\begin{figure}
	\centering
	\includegraphics[width=1.0\linewidth]{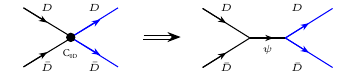}\\
    \caption{ The $D\bar{D}$ FSI treated via the resonance saturation mechanism, with $\psi$ (including $\psi(2S)$, and  $\psi(3770)$)  exchanged in the $s$ channel.}
    \label{fig:vertexc}
\end{figure}

In general, we can replace the contact coupling $C_{0D}$ in Eq.~(\ref{eq:XEFTeL}) with the $P$-wave scattering amplitude $T_{D\bar{D}}$ to account for the FSI effects~\cite{Guo:2019twa,Dai:2019hrf,Jia:2023hvc},
\begin{eqnarray}\label{eq:DDDDCP}
C_{0D}\to T_{D\bar{D}}=\frac{2\pi}{\mu_{DD}} \frac{p^2}{1/a+ip^3}\,.
\end{eqnarray}
However, since the P-wave $D\bar{D}$ interaction lacks a well-determined scattering length, we account for this effect via the resonance saturation mechanism. Specifically, the vector charmonium resonances $\psi(2S)$ and $\psi(3770)$ are introduced to saturate the scattering amplitude, realizing the $D\bar{D}$ FSI via an $s$-channel propagator as shown in Fig.~\ref{fig:vertexc}. Under the heavy quark limit, the effective Lagrangians for the interactions of $\psi(2S)$ and $\psi(3770)$ with charmed mesons can be written as \cite{Casalbuoni:1996pg,Colangelo:2003sa,Guo:2013zbw}
\begin{eqnarray}\label{eq:psiDD}
 \mathcal{L}&=&ig_{\psi(2S) D \bar{D}}\psi^i \bar{D}^{\dagger} \pararrowk{i}  D^{\dagger} \nonumber \\ && + ig_{\psi(3770) D \bar{D}}\psi^i \bar{D}^{\dagger} \pararrowk{i}  D^{\dagger} \, ,
\end{eqnarray}
where the coupling constant $g_{\psi(2S) D \bar{D}}=\sqrt{ m_{\psi(2S)}}/(2\sqrt{2} m_D f_{\psi(2S)})$~\cite{Colangelo:2003sa,Guo:2010ak}. The $\psi(2S)$ decay constant $f_{\psi(2S)}$ can be extracted from the dielectron decay width $\Gamma(\psi(2S) \to e^+e^-)$~\cite{Shi:2023ntq,Liu:2024hba,Yuan:2025pnt}. Using the PDG data~\cite{ParticleDataGroup:2024cfk}, we obtain $f_{\psi(2S)}=294~\rm MeV$. The coupling constant $g_{\psi(3770) D \bar{D}}$ is extracted from the experimental branching ratios of $\psi(3770) \to D^0{\bar D}^0$ and $D^+D^-$~\cite{ParticleDataGroup:2024cfk}, which gives $g_{\psi(3770) D^0{\bar D}^0} = 1.24 \, \rm GeV^{-\frac{3}{2}}$ and $g_{\psi(3770) D^+D^-} =1.32\, \rm GeV^{-\frac{3}{2}}$~\cite{Li:2013zcr}.

The Coulomb interaction is neglected because our primary objective is to investigate strong FSI effects in both the $X(3872)\to D^0\bar{D}^0\gamma$ and $X(3872)\to D^+D^-\gamma$ channels. The Sommerfeld enhancement arising from the Coulomb potential is a purely electromagnetic correction. Such an enhancement only exists for the charged $D^+D^-$ pair, while the neutral $D^0 \bar{D}^0$ channel carries no analogous electromagnetic effect. This electromagnetic correction is not the focus of the present work and thus lies outside the scope of our main model.

\subsection{Amplitudes of $X(3872)\to D \bar{D}\gamma$}
\label{subsec:amp-XDD}

Based on the Lagrangians above, we can write down the amplitudes of $X(3872)\to D \bar{D}\gamma$ shown in Fig.~\ref{fig:feyndiags}. The Breit-Wigner form of the $D^*$ propagator, $G_{D^*}(p)$, is used to include the contribution of the $D^*$ self-energy, i.e.,
\begin{align}\label{eq:propagator}
    G_{D^*}(p)=
    \frac{i}{p^{0}_{D^*}-m_{D^*}-\frac{\vec{p}_{D^*}^{\,2}}{2m_{D^*}}+i\frac{\Gamma_{D^*}}{2}}\,,
\end{align}
where $p=(p^0_{D^*},\vec{p}_{D^*})$ is the four-momentum of the $D^*$, $\Gamma_{D^{*+}}=83.4$ keV~\cite{ParticleDataGroup:2024cfk} and  $\Gamma_{D^{*0}}=55.3$ keV~\cite{Guo:2019qcn}.

The amplitude of the decay $X(3872) \to  D^0 \bar{D}^0\gamma$ from the tree-level diagram in  Fig.~\ref{fig:feyndiags}(a) is
\begin{align}
    \label{eq:ampa}
  i\mathcal{A}_a &= \frac{-g_n  \mu_n\mu_{D^0}}{\sqrt{2}(\gamma_{n}^2+\Vec{p}_2^{\,2}-i\Gamma_{D^{*0}} \mu_n)}\varepsilon^{ijk} q^{i}_{\gamma} \epsilon^{j*}(\gamma) \epsilon^k(X)  \,.
\end{align}
Here, $\vec{q}_{\gamma}$ is the three momentum of the final state $\gamma$, $ \epsilon^k(X)$ and $\epsilon^{j*}(\gamma)$ are the polarization vectors of the initial state particle $X(3872)$ and the final state particle $\gamma$, respectively, and $\Vec{p}_2$ is the three-momentum of the final state $\bar{D}^0$.

The amplitudes for the decay $X(3872) \to \gamma \psi \to  D^0 \bar{D}^0\gamma$ from triangle loop diagrams in  Figs.~\ref{fig:feyndiags}(b) and (c) are
\begin{align}\label{eq:ampb}
i\mathcal{A}_{b} =&\frac{- g_n \mu_{D^0} g_{\psi n}^2(\Vec{p}_1^{\,2}-\Vec{p}_2^{\,2})} {2\sqrt{2}(p^0_{\psi}-m_{\psi}-\Vec{q}_{\gamma}^{\,2}/(2m_\psi)+i\Gamma_{\psi}/2)} \nonumber \\ \times& \left[2I^{(1)}_b(q_{\gamma})-I_b(q_\gamma)\right]\varepsilon^{ijk} q^i_{\gamma} \epsilon^{j*}(\gamma) \epsilon^k(X) \,,
\end{align}
\begin{align}\label{eq:ampc}
   i\mathcal{A}_{c} =&\frac{- g_c \mu_{D^+} g_{\psi n} g_{\psi c}(\Vec{p}_1^{\,2}-\Vec{p}_2^{\,2})} {2\sqrt{2}(p^0_{\psi}-m_{\psi}-\Vec{q}_{\gamma}^{\,2}/(2m_\psi)+i\Gamma_{\psi}/2)} \nonumber \\ \times&\left[2I^{(1)}_c(q_{\gamma})-I_c(q_\gamma)\right ]\varepsilon^{ijk} q^i_{\gamma} \epsilon^{j*}(\gamma) \epsilon^k(X) \,,
\end{align}
where $g_{\psi n}$ and $g_{\psi c}$ are the coupling constants for the $\psi$ with $D^0\bar{D}^0$ and $D^+D^-$ interaction, respectively, and $m_{\psi}$ and $\Gamma_{\psi}$ are the mass and width of $\psi$, respectively. Here, $\Vec{p}_1$ and $\Vec{p}_2$ are the three momenta of the two final states $D$ and $\bar{D}$, respectively. The three-point scalar loop integrals $I(q_{\gamma})$ and vector loop integrals $I^{(1)}(q_{\gamma})$~\cite{Guo:2010ak,Dai:2019hrf,Jia:2023hvc} are given in Appendix~\ref{sec:Triangle loop}. According to Fig.~\ref{fig:feyndiags}, $m_1$, $m_2$ and $m_3$ in Eqs.~(\ref{Eq:loop_integral}) and (\ref{Eq:vector loop integral}) are identified with the masses of $D^{*0}$, $\bar{D}^0$, and $D^0$ for $I_b(q_\gamma)$ and $I^{(1)}_b(q_\gamma)$ and with those of $D^{*+}$, $D^-$, and $D^+$ for $I_c(q_\gamma)$ and $I^{(1)}_c(q_\gamma)$, respectively.

The amplitude of the decay $X(3872) \to D^+ D^-\gamma$ from the tree-level diagram in  Fig.~\ref{fig:feyndiags}(d) is
\begin{align}\label{eq:ampd}
  i\mathcal{A}_d &= \frac{-g_c \mu_c\mu_{D^+}}{\sqrt{2}(\gamma_{c}^2+\Vec{p}_2^{\,2}-i\Gamma_{D^{*+}} \mu_c)} \varepsilon^{ijk} q^i_{\gamma} \epsilon^{j*}(\gamma) \epsilon^k(X)  \,.
\end{align}
Here, $\Vec{p}_2$ is the three-momentum of the external $D^{-}$.

The amplitudes of the decay $X(3872)\to \gamma \psi \to D^+  D^- \gamma$ from triangle loop diagrams in Figs.~\ref{fig:feyndiags}(e) and \ref{fig:feyndiags}(f) are

\begin{align}\label{eq:ampe}
i\mathcal{A}_{e} =&\frac{- g_c \mu_{D^+} g_{\psi c}^2(\Vec{p}_1^{\,2}-\Vec{p}_2^{\,2})} {2\sqrt{2}(p^0_{\psi}-m_{\psi}-\Vec{q}_{\gamma}^{\,2}/(2m_\psi)+i\Gamma_{\psi}/2)} \nonumber \\ \times&\left[2I^{(1)}_c(q_{\gamma})-I_c(q_\gamma)\right ]\varepsilon^{ijk} q^i_{\gamma} \epsilon^{j*}(\gamma) \epsilon^k(X) \,,
\end{align}
\begin{align}\label{eq:ampf}
i\mathcal{A}_{f} =&\frac{- g_n \mu_{D^0}g_{\psi n} g_{\psi c}(\Vec{p}_1^{\,2}-\Vec{p}_2^{\,2})} {2\sqrt{2}(p^0_{\psi}-m_{\psi}-\Vec{q}_{\gamma}^{\,2}/(2m_\psi)+i\Gamma_{\psi}/2)} \nonumber \\ \times&\left[2I^{(1)}_b(q_{\gamma})-I_b(q_\gamma)\right ]\varepsilon^{ijk} q^i_{\gamma} \epsilon^{j*}(\gamma) \epsilon^k(X) \,.
\end{align}
As shown in these equations, the $D\bar{D}$ FSIs are described by the explicit triangle loop diagrams with $\psi(2S)$ and $\psi(3770)$ propagators. The momentum dependence is fully and consistently integrated over in the loop functions $I(q_\gamma)$ and $I^{(1)}(q_\gamma)$ via the Breit-Wigner propagator of $\psi$.

The partial decay width for $X(3872)\to D \bar{D} \gamma$ is
\begin{eqnarray}\label{eq:dwr}
    d\Gamma=2m_{X}2E_1 2E_2 \frac{1}{6m_{X}}\sum_{\rm spins}\left\vert {\mathcal{A}_{tot}}\right\vert^2 d\Phi_3\,,
\end{eqnarray}
where the $2m_{X}2E_1 2E_2$ is the normalization factor for nonrelativistic fields, and the symbol $\sum_{\rm spins}$ denotes a sum over the polarizations of the initial state $X(3872)$ and the final state $\gamma$. For the neutral channel $X(3872)\to  D^0 \bar{D}^0\gamma$, the total amplitude $\mathcal{A}_{tot}$ is the sum of the tree-level and triangle loops contributions, including the charge-conjugated diagrams,
\begin{eqnarray}
\mathcal{A}^{n}_{tot} = \mathcal{A}_a + \mathcal{A}_b + \mathcal{A}_c + \mathcal{A}(\text{c.c.})\,.
\end{eqnarray}
For the charged channel $X(3872)\to D^+ D^- \gamma$, the corresponding total amplitude $\mathcal{A}_{tot}$ reads
\begin{eqnarray}
\mathcal{A}^{c}_{tot} = \mathcal{A}_d + \mathcal{A}_e + \mathcal{A}_f + \mathcal{A}(\text{c.c.})\,.
\end{eqnarray}
The $d\Phi_3$ is the three body phase space integration
\begin{eqnarray}\label{eq:threebodyps}
 \int d\Phi_3=\int\frac{1}{32\pi^3}\frac{1}{4E_1 E_2}d \vert \Vec{p}_1 \vert^2 d \vert \Vec{p}_2 \vert^2 \,,
\end{eqnarray}
where $\Vec{p}_1$ and $\Vec{p}_2$ are the three momenta of the two final states $D$ and $\bar{D}$, respectively.

\section{Numerical Results}\label{sec:results}

In this section, we present the partial decay widths of $X(3872)\to D^0 \bar{D}^0  \gamma$ and $X(3872)\to D^+  D^-\gamma$. Here, we fix the $X(3872)$ mass at $m_{X}=3871.64 ~\rm MeV$ and adopt the phase angle $\theta=\pi/4$, which corresponds to equal weights for the neutral and charged components in the $X(3872)$ wave function. Table~\ref{tab:results} summarizes the partial decay widths (in units of keV) obtained from tree-level diagrams ($\Gamma_{\rm Tree}$), triangle loop diagrams ($\Gamma_{\rm Tri}$) and from the full calculation that includes both the tree-level and the triangle loop diagrams contributions ($\Gamma_{\rm Total}$). For the $X(3872)\to D^0 \bar{D}^0 \gamma $ decay process, the contribution of the triangle loop diagrams to the total decay width is very small, reducing it by $0.2\%$ relative to the tree-level result. However, for the $X(3872)\to D^+ D^-\gamma $ decay, the contribution of the triangle loop diagrams is relatively more pronounced, increasing the total decay width by $1.3\%$ relative to the tree-level result. Both total widths remain well below the corresponding experimental upper limits~\cite{ParticleDataGroup:2024cfk}.

\begin{table*}[htbp]
\caption{Partial decay widths (in units of keV) of the $X(3872)$ with $m_{X}=3871.64 ~\rm MeV$ and $\theta=\pi/4$. $\Gamma_{\rm Tree}$ denotes results from  tree-level diagrams, $\Gamma_{\rm Tri}$ denotes results from triangle loop diagrams, and $\Gamma_{\rm Total}$ includes both the tree-level and triangle loops diagrams contributions. The last column lists the experimental upper limits~\cite{ParticleDataGroup:2024cfk}.}
	\label{tab:results}
	\begin{ruledtabular}
		\begin{tabular}{lccccc}
		Decay channel	&$\Gamma_{\rm Tree}$ &$\Gamma_{\rm Tri}$&$\Gamma_{\rm Total}$ & Exp.data\\
       $X(3872)\to D^0\bar{D}^0\gamma$         &$7.26$&  $4.91 \times 10^{-4}$    &$7.24$& $<83.30$    \\
      $X(3872)\to D^+ D^- \gamma$    &$2.09 \times 10^{-2}$ &  $2.70\times 10^{-5} $ &$2.12\times 10^{-2}$ & $<47.60$     \\
		\end{tabular}
	\end{ruledtabular}
\end{table*}

\begin{figure}
	\centering
	\includegraphics[width=0.95\linewidth]{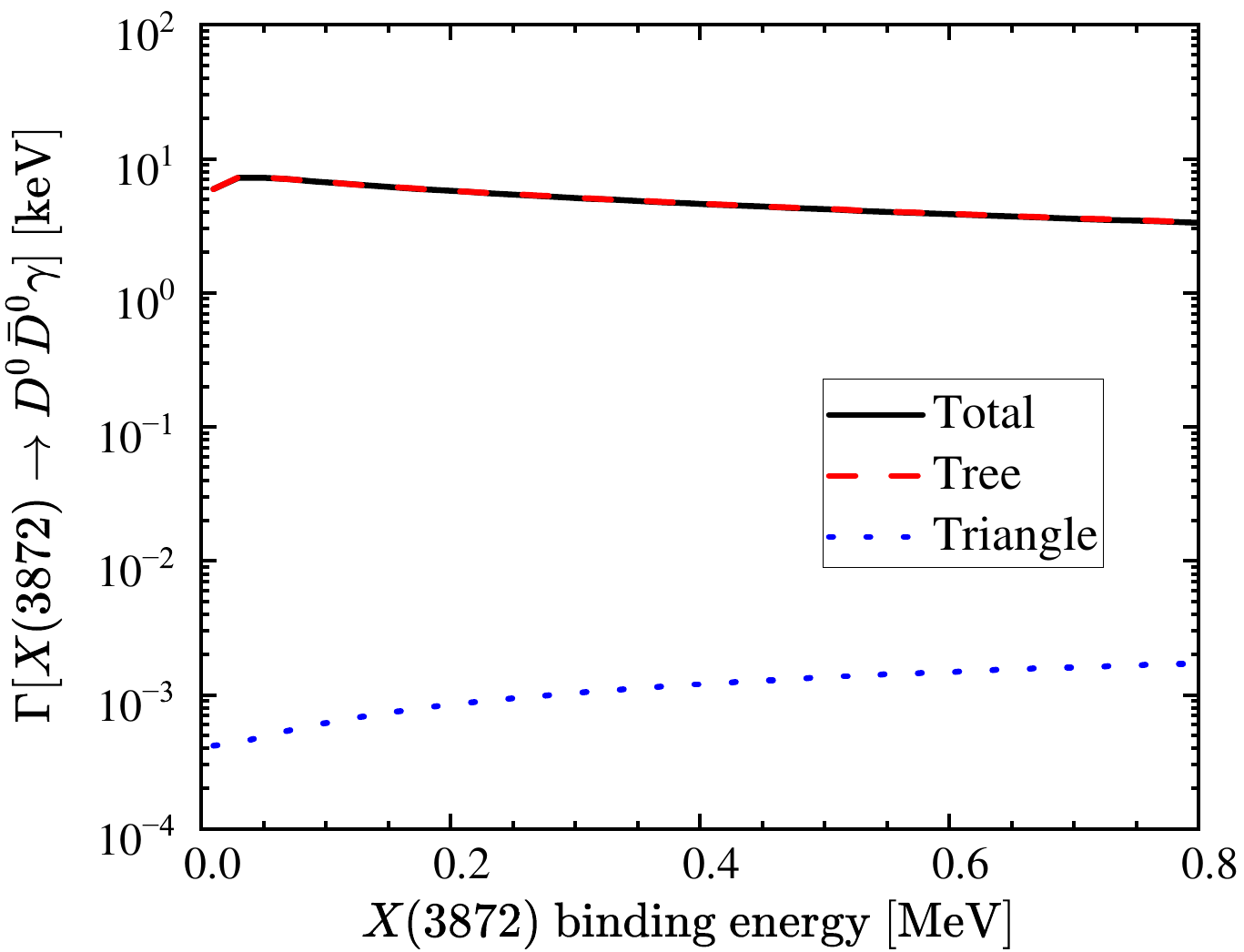}
	\caption{Decay widths for the process $X(3872) \to  D^0 \bar{D}^0\gamma$ as a function of the binding energy of $X(3872)$. }
	\label{fig:widthEn1}   
\end{figure}
\begin{figure}
	\centering
	\includegraphics[width=0.95\linewidth]{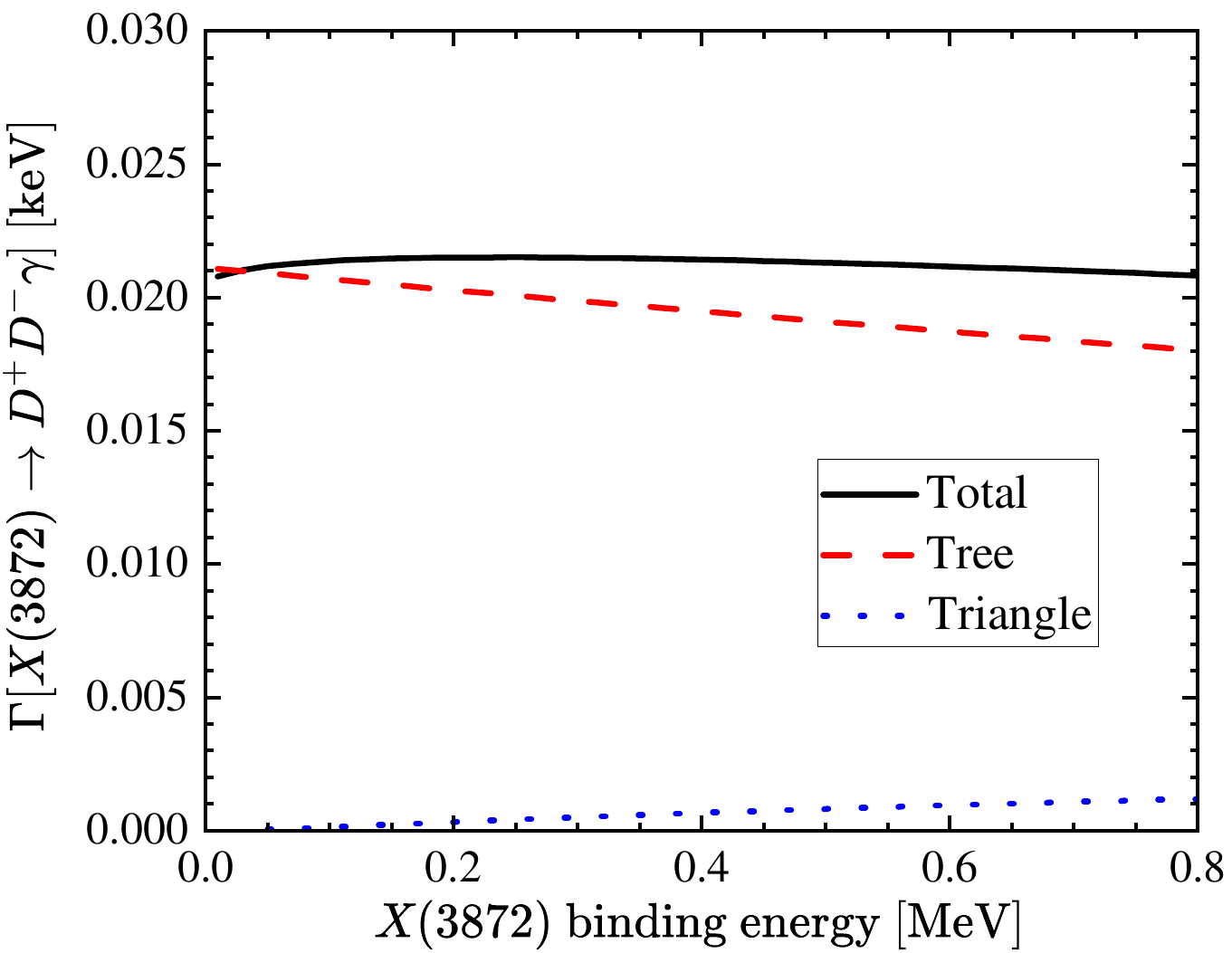}
	\caption{Decay widths for the process $X(3872) \to  D^+ D^-\gamma$ as a function of the binding energy of $X(3872)$. }
	\label{fig:widthEn2}   
\end{figure}
To clearly see the dependence of the decay widths on the binding energy $E_n$, we vary $E_n$ in the range $0.01\text{--}0.80$ MeV and plot the corresponding widths at a phase angle of $\theta=\pi/4$. Figure~\ref{fig:widthEn1} shows the results for $X(3872)\to D^0 \bar{D}^0 \gamma$. The red dashed line denotes the tree-level decay width, which exhibits a nonmonotonic behavior, first increasing and then decreasing as $E_n$ increases, and can reach about $7.3$ keV at smaller binding energies. The blue dotted line represents the contribution from the triangle loop diagrams, which grows monotonically with increasing $E_n$ and is less than $1$ keV. The black solid line corresponds to the total decay width incorporating both tree-level and triangle loop diagrams contributions, which follows the same nonmonotonic trend as the tree-level result and peaks at about $7.3$ keV. For binding energies ranging from $0.01$ MeV to $0.80$ MeV, the results obtained from tree-level diagrams and the full calculation are substantially larger than those from triangle loop diagrams, and the triangle loop diagrams yield a maximum contribution of about $1\%$ in the total width compared with the tree-level result.

For the decay $X(3872)\to D^+ D^-\gamma$, as shown in Fig.~\ref{fig:widthEn2}, all the decay widths remain below $1.0\,\mathrm{keV}$ over the entire $E_n$ range considered. The tree-level decay widths decrease as $E_n$ increases, while the total decay widths first increase and then decrease as $E_n$ increases. Over most of the binding-energy range considered in this work, the partial decay width from all diagrams is the largest, followed by the tree-level decay width, while the decay width from triangle loop diagrams is the smallest. For binding energies ranging from $0.01$ MeV to $0.80$ MeV, the maximum contribution from the triangle loop diagrams enhances the total decay width by approximately $16\%$ relative to the tree-level width.

\begin{figure}
	\centering
	\includegraphics[width=0.95\linewidth]{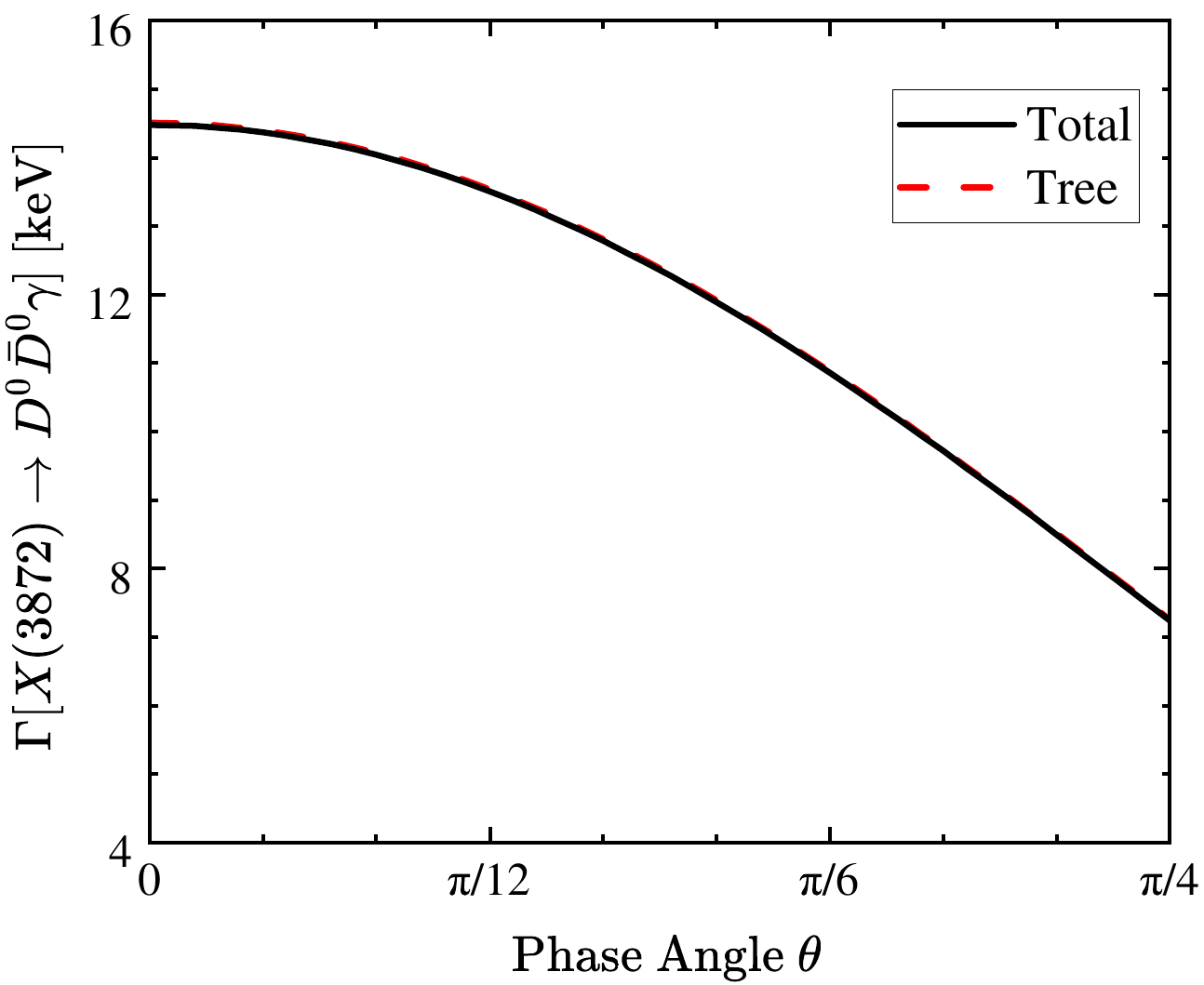}
	\caption{Decay widths for the process $X(3872) \to  D^0 \bar{D}^0\gamma$ as a function of the phase angle. }
	\label{fig:widthntheta}   
\end{figure}
\begin{figure}
	\centering
	\includegraphics[width=0.95\linewidth]{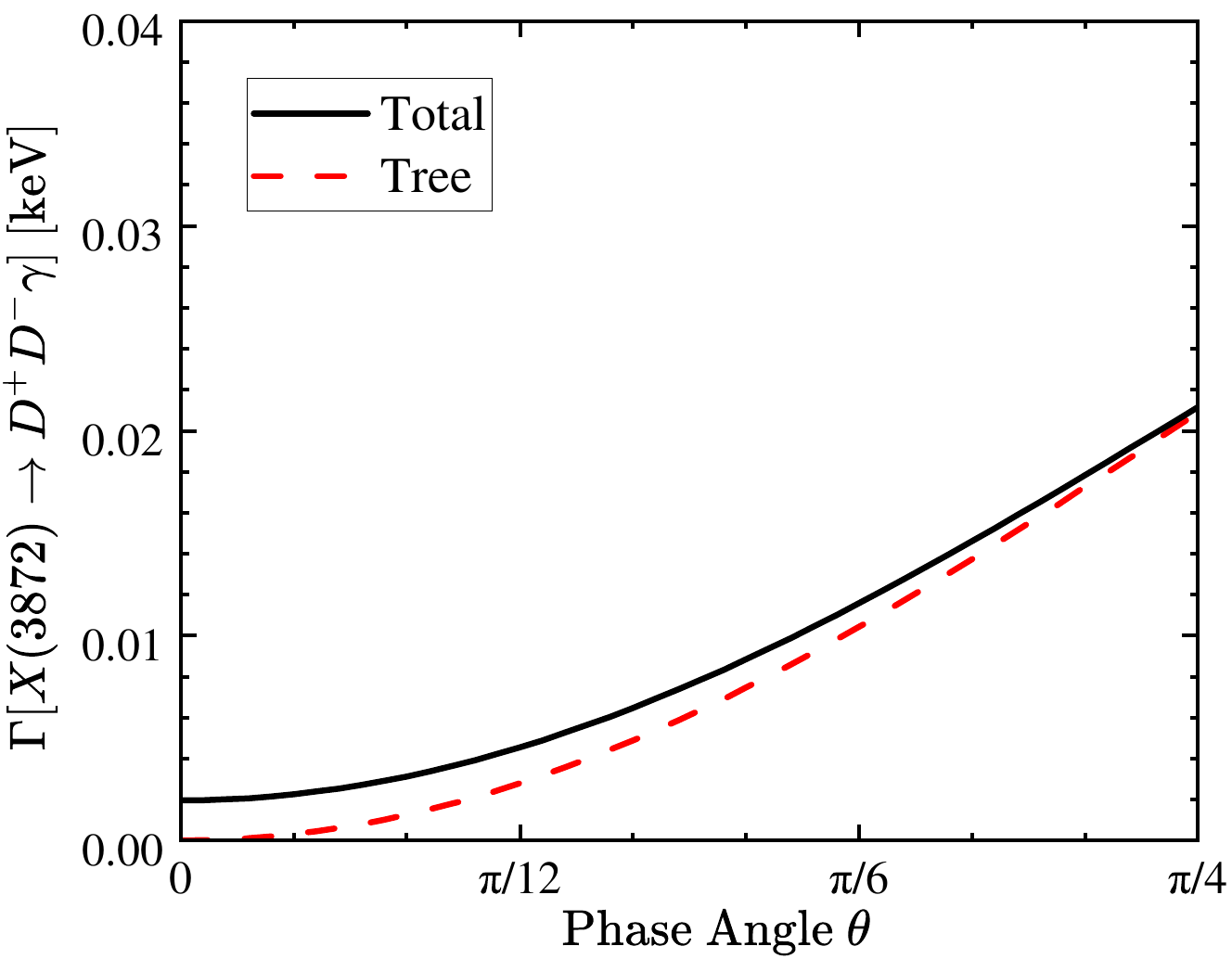}
	\caption{Decay widths for the process $X(3872) \to D^+ D^- \gamma$ as a function of the phase angle. }
	\label{fig:widthctheta}   
\end{figure}

To clearly show the influence of the phase angle, we present in Figs.~\ref{fig:widthntheta} and \ref{fig:widthctheta} the partial decay widths for $X(3872)\to D^0\bar{D}^0 \gamma$ and $X(3872)\to D^+ D^-\gamma$ as functions of the phase angle $\theta$ with $m_X=3871.64\,\mathrm{MeV}$. For the decay $X(3872) \to  D^0 \bar{D}^0\gamma$, both the tree-level and total partial decay widths decrease monotonically as $\theta$ increases from $0$ to $\pi/4$. In contrast, for the $X(3872) \to D^+ D^-\gamma$, the tree-level width and total partial decay width increase monotonically with $\theta$. For the decay $X(3872) \to D^+ D^- \gamma$, as the phase angle increases from $0$ to $\pi/4$, the difference between the tree-level decay width and the total decay width gradually decreases.

\section{Summary}\label{section:summary}

Within nonrelativistic effective field theory, we have studied the radiative decays $X(3872)\to  D^0 \bar{D}^0\gamma$ and $X(3872)\to  D^+ D^-\gamma$ assuming $X(3872)$ to be a $D\bar{D}^*+\mathrm{c.c.}$ molecular state with $J^{PC}=1^{++}$. Our calculation includes both tree-level and triangle loop diagrams contributions, with the neutral and charged components equally mixed ($\theta=\pi/4$). At this mixing angle, the tree-level partial width for $D^0 \bar{D}^0\gamma$ is about $7.3$ keV, and the triangle loop diagrams produce a modest reduction of less than $1\%$ in the total width. In contrast, the charged channel tree-level partial width is less than $1.0$ keV, but the triangle loops cause a relative enhancement of about $1.3\%$ in the corresponding total width. The hierarchy between the two channels stems primarily from the much smaller $D^{*+}\to D^+\gamma$ magnetic transition compared to that of $D^{*0}\to D^0\gamma$. The dependence of the total decay width on the binding energy $E_n$ shows similar behaviors. For the neutral channel, the total decay width first increases and then decreases with $E_n$ increasing and can reach $7.3$ keV, and the charged-channel width shows a similar nonmonotonic behavior, but the decay width is less than $1.0$ keV. With respect to the phase angle $\theta$, the partial widths decrease monotonically for the neutral channel, while the total partial width exhibits a distinct behavior that increases monotonically for the charged channel. All predicted partial widths lie well below the present experimental upper limits. The strong hierarchy between the neutral and charged decay modes provides a clear, testable signature of the molecular picture, which can be verified in future experiments.

\begin{acknowledgments}
\label{sec:acknowledgements}
This work is supported by the National Natural Science Foundation of China under Grant No. 12475081; by the Natural Science Foundation of Shandong Province under Grant No. ZR2025MS04 and No. ZR2022ZD26; and by Taishan Scholar Project of Shandong Province under Grant No. tsqn202607074.
\end{acknowledgments}

\begin{appendix}
\section{Three-point loop integrals}\label{sec:Triangle loop}

Here, we present the loop integrals used in the calculation, where the scalar and vector three-point loop integral are ultraviolet convergent~\cite{Guo:2010ak,Jia:2023hvc}
\onecolumngrid
\begin{align}\label{Eq:loop_integral}
I(q)&=i\int \frac{d^4l}{(2 \pi)^{4}} \frac{1}{\left[l^0-m_{1}-\frac{\vec{l}^{\,2}}{2 m_{1}}+i \epsilon\right]\left[M-l^{0}-m_{2}-\frac{\vec{l}^{\,2}}{2 m_{2}}+i \epsilon\right]\left[l^{0}-q^0-m_{3}-\frac{\left(\vec{l}-\vec{q}\right)^{2}}{2 m_{3}}+i \epsilon\right]}\nonumber \\
&=\int \frac{d^{3}l}{(2 \pi)^{3}} \frac{1}{\left[b_{12}+\frac{\vec{l}^{\,2}}{2 \mu_{12}}-i\epsilon\right]\left[b_{23}+\frac{\vec{l}^{\,2}}{2 m_{2}}+\frac{\left(\vec{l}-\vec{q}\right)^{2}}{2 m_{3}}-i \epsilon\right]}\nonumber \\
&=\frac{\mu_{12} \mu_{23}}{2 \pi } \frac{1}{\sqrt{a}}\left[\tan ^{-1}\left(\frac{c_2-c_1}{2 \sqrt{a c_1}}\right)+\tan ^{-1}\left(\frac{2 a+c_1-c_2}{2 \sqrt{a\left(c_2-a\right)}}\right)\right]\, ,
\end{align}
where $\mu_{ij}=m_{i} m_{j} /\left(m_{i}+m_{j}\right)$ are the reduced masses, $b_{12} = m_1+m_2-M$, $b_{23} = m_{2}+m_{3}+q^0-M$, and the parameters
\begin{align}
a = \left(\frac{\mu_{23}}{m_3}\right)^2 \vec{q}^{\, 2}, \quad c_1 = 2 \mu_{12} b_{12}, \quad
c_2 = 2 \mu_{23} b_{23}+\frac{\mu_{23}}{m_3} \vec{q}^{\,2}\, .
\end{align}
The vector integral is defined as
\begin{align}
q^{i} I^{(1)}(q)&=\int \frac{d^{3} \vec{l}}{(2 \pi)^{3}}  \frac{l^{i}}{\left(b_{12}+\frac{\vec{l}^{2}}{2 \mu_{12}}-i \epsilon\right)\left[b_{23}+\frac{\vec{l}^{2}}{2 m_{2}}+\frac{\left(\vec{l}-\vec{q}\right)^{2}}{2 m_{3}}-i \epsilon\right]} \nonumber \\
 &=\frac{\mu_{23}}{a m_3}\left[B\left(c_2-a\right)-B\left(c_1\right)+\frac{1}{2}\left(c_2-c_1\right) I(q)\right] \, ,
 \label{Eq:vector loop integral}
\end{align}
where the two-point function $B(c)$ in the power divergence subtraction (PDS) scheme~\cite{Kaplan:1998tg} reads
\begin{align}
    B(c)&=2 \mu_{12} \mu_{23} \left(\frac{\Lambda_{\mathrm{PDS}}}{2}\right)^{4-d} \int \frac{d^{d-1} l}{(2 \pi)^{d-1}} \frac{1}{\vec{l}^{\,2}+c-i \epsilon} \nonumber \\
&=\frac{\mu_{12} \mu_{23}}{2 \pi}\left(\Lambda_{\mathrm{PDS}}-\sqrt{c-i \epsilon}\right),
\end{align}
with $\Lambda_{\rm{PDS}}$ a scale in the PDS scheme.
\twocolumngrid
\end{appendix}

\bibliography{X3872decay}

@article{Richard:2016eis,
    author = "Richard, Jean-Marc",
    title = "{Exotic hadrons: review and perspectives}",
    eprint = "1606.08593",
    archivePrefix = "arXiv",
    primaryClass = "hep-ph",
    doi = "10.1007/s00601-016-1159-0",
    journal = "Few Body Syst.",
    volume = "57",
    number = "12",
    pages = "1185--1212",
    year = "2016"
}

@article{Hosaka:2016pey,
    author = "Hosaka, Atsushi and Iijima, Toru and Miyabayashi, Kenkichi and Sakai, Yoshihide and Yasui, Shigehiro",
    title = "{Exotic hadrons with heavy flavors: $X, Y, Z$, and related states}",
    eprint = "1603.09229",
    archivePrefix = "arXiv",
    primaryClass = "hep-ph",
    reportNumber = "J-PARC-TH-0046",
    doi = "10.1093/ptep/ptw045",
    journal = "PTEP",
    volume = "2016",
    number = "6",
    pages = "062C01",
    year = "2016"
}

@article{Guo:2017jvc,
    author = "Guo, Feng-Kun and Hanhart, Christoph and Mei{\ss}ner, Ulf-G. and Wang, Qian and Zhao, Qiang and Zou, Bing-Song",
    title = "{Hadronic molecules}",
    eprint = "1705.00141",
    archivePrefix = "arXiv",
    primaryClass = "hep-ph",
    doi = "10.1103/RevModPhys.90.015004",
    journal = "Rev. Mod. Phys.",
    volume = "90",
    number = "1",
    pages = "015004",
    year = "2018",
    note = "[Erratum: Rev.Mod.Phys. 94, 029901 (2022)]"
}

@article{Ali:2017jda,
    author = {Ali, Ahmed and Lange, Jens S{\"o}ren and Stone, Sheldon},
    title = "{Exotics: Heavy Pentaquarks and Tetraquarks}",
    eprint = "1706.00610",
    archivePrefix = "arXiv",
    primaryClass = "hep-ph",
    reportNumber = "DESY-17-071",
    doi = "10.1016/j.ppnp.2017.08.003",
    journal = "Prog. Part. Nucl. Phys.",
    volume = "97",
    pages = "123--198",
    year = "2017"
}

@article{Liu:2019zoy,
    author = "Liu, Yan-Rui and Chen, Hua-Xing and Chen, Wei and Liu, Xiang and Zhu, Shi-Lin",
    title = "{Pentaquark and Tetraquark states}",
    eprint = "1903.11976",
    archivePrefix = "arXiv",
    primaryClass = "hep-ph",
    doi = "10.1016/j.ppnp.2019.04.003",
    journal = "Prog. Part. Nucl. Phys.",
    volume = "107",
    pages = "237--320",
    year = "2019"
}

@article{Belle:2003nnu,
    author = "Choi, S. K. and others",
    collaboration = "Belle",
    title = "{Observation of a narrow charmonium-like state in exclusive $B^\pm \to K^\pm \pi^+ \pi^- J/\psi$ decays}",
    eprint = "hep-ex/0309032",
    archivePrefix = "arXiv",
    doi = "10.1103/PhysRevLett.91.262001",
    journal = "Phys. Rev. Lett.",
    volume = "91",
    pages = "262001",
    year = "2003"
}

@article{LHCb:2013kgk,
    author = "Aaij, R and others",
    collaboration = "LHCb",
    title = "{Determination of the $X(3872)$ meson quantum numbers}",
    eprint = "1302.6269",
    archivePrefix = "arXiv",
    primaryClass = "hep-ex",
    reportNumber = "LHCB-PAPER-2013-001, CERN-PH-EP-2013-017",
    doi = "10.1103/PhysRevLett.110.222001",
    journal = "Phys. Rev. Lett.",
    volume = "110",
    pages = "222001",
    year = "2013"
}

@article{Chen:2022asf,
    author = "Chen, Hua-Xing and Chen, Wei and Liu, Xiang and Liu, Yan-Rui and Zhu, Shi-Lin",
    title = "{An updated review of the new hadron states}",
    eprint = "2204.02649",
    archivePrefix = "arXiv",
    primaryClass = "hep-ph",
    doi = "10.1088/1361-6633/aca3b6",
    journal = "Rept. Prog. Phys.",
    volume = "86",
    number = "2",
    pages = "026201",
    year = "2023"
}

@article{ParticleDataGroup:2024cfk,
    author = "Navas, S. and others",
    collaboration = "Particle Data Group",
    title = "{Review of particle physics}",
    doi = "10.1103/PhysRevD.110.030001",
    journal = "Phys. Rev. D",
    volume = "110",
    number = "3",
    pages = "030001",
    year = "2024"
}

@article{Close:2003sg,
    author = "Close, Frank E. and Page, Philip R.",
    title = "{The $D^{*0}\bar{D}^0$  threshold resonance}",
    eprint = "hep-ph/0309253",
    archivePrefix = "arXiv",
    doi = "10.1016/j.physletb.2003.10.032",
    journal = "Phys. Lett. B",
    volume = "578",
    pages = "119--123",
    year = "2004"
}

@article{Pakvasa:2003ea,
    author = "Pakvasa, Sandip and Suzuki, Mahiko",
    title = "{On the hidden charm state at $3872$ MeV}",
    eprint = "hep-ph/0309294",
    archivePrefix = "arXiv",
    reportNumber = "UH-511-1034-03, LBNL-53823",
    doi = "10.1016/j.physletb.2003.11.005",
    journal = "Phys. Lett. B",
    volume = "579",
    pages = "67--73",
    year = "2004"
}

@article{Swanson:2004pp,
    author = "Swanson, Eric S.",
    title = "{Diagnostic decays of the $X(3872)$}",
    eprint = "hep-ph/0406080",
    archivePrefix = "arXiv",
    doi = "10.1016/j.physletb.2004.07.059",
    journal = "Phys. Lett. B",
    volume = "598",
    pages = "197--202",
    year = "2004"
}

@article{Swanson:2003tb,
    author = "Swanson, Eric S.",
    title = "{Short range structure in the $X(3872)$}",
    eprint = "hep-ph/0311229",
    archivePrefix = "arXiv",
    reportNumber = "JLAB-THY-03-227",
    doi = "10.1016/j.physletb.2004.03.033",
    journal = "Phys. Lett. B",
    volume = "588",
    pages = "189--195",
    year = "2004"
}

@article{Tornqvist:2004qy,
    author = "Tornqvist, Nils A.",
    title = "{Isospin breaking of the narrow charmonium state of Belle at $3872$ MeV as a deuson}",
    eprint = "hep-ph/0402237",
    archivePrefix = "arXiv",
    doi = "10.1016/j.physletb.2004.03.077",
    journal = "Phys. Lett. B",
    volume = "590",
    pages = "209--215",
    year = "2004"
}

@article{Voloshin:2003nt,
    author = "Voloshin, M. B.",
    title = "{Interference and binding effects in decays of possible molecular component of $X(3872)$}",
    eprint = "hep-ph/0309307",
    archivePrefix = "arXiv",
    reportNumber = "TPI-MINN-03-27-T, UMN-TH-2216-03",
    doi = "10.1016/j.physletb.2003.11.014",
    journal = "Phys. Lett. B",
    volume = "579",
    pages = "316--320",
    year = "2004"
}

@article{Wong:2003xk,
    author = "Wong, Cheuk-Yin",
    title = "{Molecular states of heavy quark mesons}",
    eprint = "hep-ph/0311088",
    archivePrefix = "arXiv",
    doi = "10.1103/PhysRevC.69.055202",
    journal = "Phys. Rev. C",
    volume = "69",
    pages = "055202",
    year = "2004"
}

@article{AlFiky:2005jd,
    author = "AlFiky, Mohammad T. and Gabbiani, Fabrizio and Petrov, Alexey A.",
    title = "{$X(3872)$: Hadronic molecules in effective field theory}",
    eprint = "hep-ph/0506141",
    archivePrefix = "arXiv",
    reportNumber = "WSU-HEP-0502",
    doi = "10.1016/j.physletb.2006.07.069",
    journal = "Phys. Lett. B",
    volume = "640",
    pages = "238--245",
    year = "2006"
}

@article{Braaten:2006sy,
    author = "Braaten, Eric and Lu, Meng",
    title = "{Operator Product Expansion in the Production and Decay of the $X(3872)$}",
    eprint = "hep-ph/0606115",
    archivePrefix = "arXiv",
    doi = "10.1103/PhysRevD.74.054020",
    journal = "Phys. Rev. D",
    volume = "74",
    pages = "054020",
    year = "2006"
}

@article{Fleming:2007rp,
    author = "Fleming, S. and Kusunoki, M. and Mehen, T. and van Kolck, U.",
    title = "{Pion interactions in the $X(3872)$}",
    eprint = "hep-ph/0703168",
    archivePrefix = "arXiv",
    reportNumber = "JLAB-THY-07-615",
    doi = "10.1103/PhysRevD.76.034006",
    journal = "Phys. Rev. D",
    volume = "76",
    pages = "034006",
    year = "2007"
}

@article{Ding:2009vj,
    author = "Ding, Gui-Jun and Liu, Jia-Feng and Yan, Mu-Lin",
    title = "{Dynamics of Hadronic Molecule in One-Boson Exchange Approach and Possible Heavy Flavor Molecules}",
    eprint = "0901.0426",
    archivePrefix = "arXiv",
    primaryClass = "hep-ph",
    doi = "10.1103/PhysRevD.79.054005",
    journal = "Phys. Rev. D",
    volume = "79",
    pages = "054005",
    year = "2009"
}

@article{Dong:2009yp,
    author = "Dong, Yubing and Faessler, Amand and Gutsche, Thomas and Kovalenko, Sergey and Lyubovitskij, Valery E.",
    title = "{$X(3872)$ as a hadronic molecule and its decays to charmonium states and pions}",
    eprint = "0903.5416",
    archivePrefix = "arXiv",
    primaryClass = "hep-ph",
    doi = "10.1103/PhysRevD.79.094013",
    journal = "Phys. Rev. D",
    volume = "79",
    pages = "094013",
    year = "2009"
}

@article{Lee:2009hy,
    author = "Lee, Ian Woo and Faessler, Amand and Gutsche, Thomas and Lyubovitskij, Valery E.",
    title = "{$X(3872)$ as a molecular $DD^*$ state in a potential model}",
    eprint = "0910.1009",
    archivePrefix = "arXiv",
    primaryClass = "hep-ph",
    doi = "10.1103/PhysRevD.80.094005",
    journal = "Phys. Rev. D",
    volume = "80",
    pages = "094005",
    year = "2009"
}

@article{Lee:2011rka,
    author = "Lee, Ning and Luo, Zhi-Gang and Chen, Xiao-Lin and Zhu, Shi-Lin",
    title = "{Possible Deuteron-like Molecular States Composed of Heavy Baryons}",
    eprint = "1104.4257",
    archivePrefix = "arXiv",
    primaryClass = "hep-ph",
    doi = "10.1103/PhysRevD.84.014031",
    journal = "Phys. Rev. D",
    volume = "84",
    pages = "014031",
    year = "2011"
}

@article{Liu:2009qhy,
    author = "Liu, Xiang and Luo, Zhi-Gang and Liu, Yan-Rui and Zhu, Shi-Lin",
    title = "{$X(3872)$ and Other Possible Heavy Molecular States}",
    eprint = "0808.0073",
    archivePrefix = "arXiv",
    primaryClass = "hep-ph",
    doi = "10.1140/epjc/s10052-009-1020-4",
    journal = "Eur. Phys. J. C",
    volume = "61",
    pages = "411--428",
    year = "2009"
}

@article{Zhang:2009vs,
    author = "Zhang, Jian-Rong and Huang, Ming-Qiu",
    title = "{$\{Q\bar{q}\}$$\{\bar{Q}^{'}q\}$ molecular states}",
    eprint = "0906.0090",
    archivePrefix = "arXiv",
    primaryClass = "hep-ph",
    doi = "10.1103/PhysRevD.80.056004",
    journal = "Phys. Rev. D",
    volume = "80",
    pages = "056004",
    year = "2009"
}

@article{Gamermann:2009uq,
    author = "Gamermann, D. and Nieves, J. and Oset, E. and Ruiz Arriola, E.",
    title = "{Couplings in coupled channels versus wave functions: application to the $X(3872)$ resonance}",
    eprint = "0911.4407",
    archivePrefix = "arXiv",
    primaryClass = "hep-ph",
    doi = "10.1103/PhysRevD.81.014029",
    journal = "Phys. Rev. D",
    volume = "81",
    pages = "014029",
    year = "2010"
}

@article{Mehen:2011ds,
    author = "Mehen, Thomas and Springer, Roxanne",
    title = "{Radiative Decays $X(3872) \to \psi(2S)\gamma$ and $\psi(4040) \to X(3872)\gamma$ in Effective Field Theory}",
    eprint = "1101.5175",
    archivePrefix = "arXiv",
    primaryClass = "hep-ph",
    doi = "10.1103/PhysRevD.83.094009",
    journal = "Phys. Rev. D",
    volume = "83",
    pages = "094009",
    year = "2011"
}

@article{Nieves:2011zz,
    author = "Nieves, J. and Valderrama, M. Pavon",
    title = "{Deriving the existence of $B\bar{B}^*$ bound states from the $X(3872)$ and Heavy Quark Symmetry}",
    eprint = "1106.0600",
    archivePrefix = "arXiv",
    primaryClass = "hep-ph",
    doi = "10.1103/PhysRevD.84.056015",
    journal = "Phys. Rev. D",
    volume = "84",
    pages = "056015",
    year = "2011"
}

@article{Nieves:2012tt,
    author = "Nieves, J. and Valderrama, M. Pavon",
    title = "{The Heavy Quark Spin Symmetry Partners of the $X(3872)$}",
    eprint = "1204.2790",
    archivePrefix = "arXiv",
    primaryClass = "hep-ph",
    doi = "10.1103/PhysRevD.86.056004",
    journal = "Phys. Rev. D",
    volume = "86",
    pages = "056004",
    year = "2012"
}

@article{Li:2012cs,
    author = "Li, Ning and Zhu, Shi-Lin",
    title = "{Isospin breaking, Coupled-channel effects and Diagnosis of $X(3872)$}",
    eprint = "1207.3954",
    archivePrefix = "arXiv",
    primaryClass = "hep-ph",
    doi = "10.1103/PhysRevD.86.074022",
    journal = "Phys. Rev. D",
    volume = "86",
    pages = "074022",
    year = "2012"
}

@article{Sun:2012sy,
    author = "Sun, Zhi-Feng and Liu, Xiang and Nielsen, Marina and Zhu, Shi-Lin",
    title = "{Hadronic molecules with both open charm and bottom}",
    eprint = "1203.1090",
    archivePrefix = "arXiv",
    primaryClass = "hep-ph",
    doi = "10.1103/PhysRevD.85.094008",
    journal = "Phys. Rev. D",
    volume = "85",
    pages = "094008",
    year = "2012"
}

@article{Guo:2013sya,
    author = "Guo, Feng-Kun and Hidalgo-Duque, Carlos and Nieves, Juan and Valderrama, Manuel Pavon",
    title = "{Consequences of Heavy Quark Symmetries for Hadronic Molecules}",
    eprint = "1303.6608",
    archivePrefix = "arXiv",
    primaryClass = "hep-ph",
    doi = "10.1103/PhysRevD.88.054007",
    journal = "Phys. Rev. D",
    volume = "88",
    pages = "054007",
    year = "2013"
}

@article{He:2014nya,
    author = "He, Jun",
    title = "{Study of the $B\bar{B}^*/D\bar{D}^*$ bound states in a Bethe-Salpeter approach}",
    eprint = "1409.8506",
    archivePrefix = "arXiv",
    primaryClass = "hep-ph",
    doi = "10.1103/PhysRevD.90.076008",
    journal = "Phys. Rev. D",
    volume = "90",
    number = "7",
    pages = "076008",
    year = "2014"
}

@article{Zhao:2014gqa,
    author = "Zhao, Lu and Ma, Li and Zhu, Shi-Lin",
    title = "{Spin-orbit force, recoil corrections, and possible $B \bar{B}^{*}$ and $D \bar{D}^{*}$  molecular states}",
    eprint = "1403.4043",
    archivePrefix = "arXiv",
    primaryClass = "hep-ph",
    doi = "10.1103/PhysRevD.89.094026",
    journal = "Phys. Rev. D",
    volume = "89",
    number = "9",
    pages = "094026",
    year = "2014"
}

@article{Guo:2014taa,
    author = "Guo, Feng-Kun and Hanhart, C. and Kalashnikova, Yu. S. and Mei\ss{}ner, Ulf-G. and Nefediev, A. V.",
    title = "{What can radiative decays of the $X(3872)$ teach us about its nature?}",
    eprint = "1410.6712",
    archivePrefix = "arXiv",
    primaryClass = "hep-ph",
    doi = "10.1016/j.physletb.2015.02.013",
    journal = "Phys. Lett. B",
    volume = "742",
    pages = "394--398",
    year = "2015"
}

@article{Guo:2014hqa,
    author = "Guo, F. K. and Hidalgo-Duque, C. and Nieves, J. and Ozpineci, Altug and Valderrama, M. P.",
    title = "{Detecting the long-distance structure of the $X(3872)$}",
    eprint = "1404.1776",
    archivePrefix = "arXiv",
    primaryClass = "hep-ph",
    doi = "10.1140/epjc/s10052-014-2885-4",
    journal = "Eur. Phys. J. C",
    volume = "74",
    number = "5",
    pages = "2885",
    year = "2014"
}

@article{Braaten:2003he,
    author = "Braaten, Eric and Kusunoki, Masaoki",
    title = "{Low-energy universality and the new charmonium resonance at $3870$ MeV}",
    eprint = "hep-ph/0311147",
    archivePrefix = "arXiv",
    doi = "10.1103/PhysRevD.69.074005",
    journal = "Phys. Rev. D",
    volume = "69",
    pages = "074005",
    year = "2004"
}

@article{Wu:2021udi,
    author = "Wu, Qi and Chen, Dian-Yong and Matsuki, Takayuki",
    title = "{A phenomenological analysis on isospin-violating decay of $X(3872)$}",
    eprint = "2102.08637",
    archivePrefix = "arXiv",
    primaryClass = "hep-ph",
    doi = "10.1140/epjc/s10052-021-08984-2",
    journal = "Eur. Phys. J. C",
    volume = "81",
    number = "2",
    pages = "193",
    year = "2021"
}

@article{Yamaguchi:2019vea,
    author = "Yamaguchi, Yasuhiro and Hosaka, Atsushi and Takeuchi, Sachiko and Takizawa, Makoto",
    title = "{Heavy hadronic molecules with pion exchange and quark core couplings: a guide for practitioners}",
    eprint = "1908.08790",
    archivePrefix = "arXiv",
    primaryClass = "hep-ph",
    reportNumber = "RIKEN-QHP-425",
    doi = "10.1088/1361-6471/ab72b0",
    journal = "J. Phys. G",
    volume = "47",
    number = "5",
    pages = "053001",
    year = "2020"
}

@article{Cai:2025inq,
    author = "Cai, Hao-Dong and Jia, Zhao-Sai and Li, Gang and Liu, Shi-Dong",
    title = "{Hidden charmed decays of $X(3872)$ within the $D \bar{D}^*$ molecular framework}",
    eprint = "2503.20183",
    archivePrefix = "arXiv",
    primaryClass = "hep-ph",
    doi = "10.1103/gn25-fc9q",
    journal = "Phys. Rev. D",
    volume = "111",
    number = "11",
    pages = "114024",
    year = "2025"
}

@article{Wang:2025zss,
    author = "Wang, Fan and Li, Gang and Liu, Shi-Dong and Wu, Qi",
    title = "{Radiative decays of $X(3872)$ in $D\bar{D}^*$ molecule scenario}",
    eprint = "2502.19887",
    archivePrefix = "arXiv",
    primaryClass = "hep-ph",
    doi = "10.1103/PhysRevD.111.094001",
    journal = "Phys. Rev. D",
    volume = "111",
    number = "9",
    pages = "094001",
    year = "2025"
}

@article{Wu:2025crk,
    author = "Wu, Qi and Sun, Zhong-Quan and Chen, Dian-Yong and Liu, Shi-Dong and Li, Gang",
    title = "{Dipion transitions from $X(3872)$ to $\chi_{cJ}\ (J=0,1,2)$}",
    eprint = "2512.21161",
    archivePrefix = "arXiv",
    primaryClass = "hep-ph",
    doi = "10.1103/19lr-96hx",
    journal = "Phys. Rev. D",
    volume = "113",
    number = "1",
    pages = "014015",
    year = "2026"
}

@article{Jia:2025xil,
    author = "Jia, Zhao-Sai and Li, Gang and Zhang, Zhen-Hua",
    title = "{Constrain the $\chi_{cJ}\to D^{(*)} \bar{D}^{(*)}$ effective couplings via the $X(3872)\to \pi^0 \chi_{cJ}$ decays}",
    eprint = "2507.16618",
    archivePrefix = "arXiv",
    primaryClass = "hep-ph",
    doi = "10.1103/51z1-qls5",
    journal = "Phys. Rev. D",
    volume = "112",
    number = "7",
    pages = "074035",
    year = "2025"
}

@article{Kalashnikova:2018vkv,
    author = "Kalashnikova, Yu S. and Nefediev, Alexey V.",
    title = "{$X(3872)$ in the molecular model}",
    eprint = "1811.01324",
    archivePrefix = "arXiv",
    primaryClass = "hep-ph",
    doi = "10.3367/UFNe.2018.08.038411",
    journal = "Phys. Usp.",
    volume = "62",
    number = "6",
    pages = "568--595",
    year = "2019"
}

@article{Maiani:2004vq,
    author = "Maiani, L. and Piccinini, F. and Polosa, A. D. and Riquer, V.",
    title = "{Diquark-antidiquarks with hidden or open charm and the nature of $X(3872)$}",
    eprint = "hep-ph/0412098",
    archivePrefix = "arXiv",
    reportNumber = "ROMA1-1396-2004, FNT-T-2004-20, BA-TH-502-04, CERN-PH-TH-2004-239",
    doi = "10.1103/PhysRevD.71.014028",
    journal = "Phys. Rev. D",
    volume = "71",
    pages = "014028",
    year = "2005"
}

@article{Maiani:2005pe,
    author = "Maiani, L. and Riquer, V. and Piccinini, F. and Polosa, A. D.",
    title = "{Four quark interpretation of $Y(4260)$}",
    eprint = "hep-ph/0507062",
    archivePrefix = "arXiv",
    doi = "10.1103/PhysRevD.72.031502",
    journal = "Phys. Rev. D",
    volume = "72",
    pages = "031502",
    year = "2005"
}

@article{Maiani:2007vr,
    author = "Maiani, L. and Polosa, A. D. and Riquer, V.",
    title = "{Indications of a Four-Quark Structure for the $X(3872)$ and $X(3876)$ Particles from Recent Belle and BABAR Data}",
    eprint = "0707.3354",
    archivePrefix = "arXiv",
    primaryClass = "hep-ph",
    doi = "10.1103/PhysRevLett.99.182003",
    journal = "Phys. Rev. Lett.",
    volume = "99",
    pages = "182003",
    year = "2007"
}

@article{Terasaki:2007uv,
    author = "Terasaki, Kunihiko",
    title = "{A New tetra-quark interpretation of $X(3872)$}",
    eprint = "0706.3944",
    archivePrefix = "arXiv",
    primaryClass = "hep-ph",
    reportNumber = "YITP-07-36, KANAZAWA-0707",
    doi = "10.1143/PTP.118.821",
    journal = "Prog. Theor. Phys.",
    volume = "118",
    pages = "821--826",
    year = "2007"
}

@article{Dubnicka:2020yxy,
    author = "Dubni{\v{c}}ka, Stanislav and Dubni{\v{c}}kov{\'a}, Anna Zuzana and Ivanov, Mikhail A. and Liptaj, Andrej",
    title = "{Dynamical Approach to Decays of $XYZ$ States}",
    doi = "10.3390/sym12060884",
    journal = "Symmetry",
    volume = "12",
    number = "6",
    pages = "884",
    year = "2020"
}

@article{Wang:2023sii,
    author = "Wang, Zhi-Gang",
    title = "{Decipher the width of the $X(3872)$ via the QCD sum rules}",
    eprint = "2310.02030",
    archivePrefix = "arXiv",
    primaryClass = "hep-ph",
    doi = "10.1103/PhysRevD.109.014017",
    journal = "Phys. Rev. D",
    volume = "109",
    number = "1",
    pages = "014017",
    year = "2024"
}

@article{Wang:2019tlw,
    author = "Wang, Zhi-Gang",
    title = "{Analysis of the hidden-charm tetraquark mass spectrum with the QCD sum rules}",
    eprint = "1908.07914",
    archivePrefix = "arXiv",
    primaryClass = "hep-ph",
    doi = "10.1103/PhysRevD.102.014018",
    journal = "Phys. Rev. D",
    volume = "102",
    number = "1",
    pages = "014018",
    year = "2020"
}

@article{Suzuki:2005ha,
    author = "Suzuki, Mahiko",
    title = "{The $X(3872)$ boson: Molecule or charmonium}",
    eprint = "hep-ph/0508258",
    archivePrefix = "arXiv",
    reportNumber = "LBL-58724",
    doi = "10.1103/PhysRevD.72.114013",
    journal = "Phys. Rev. D",
    volume = "72",
    pages = "114013",
    year = "2005"
}

@article{Barnes:2003vb,
    author = "Barnes, Ted and Godfrey, Stephen",
    title = "{Charmonium options for the $X(3872)$}",
    eprint = "hep-ph/0311162",
    archivePrefix = "arXiv",
    doi = "10.1103/PhysRevD.69.054008",
    journal = "Phys. Rev. D",
    volume = "69",
    pages = "054008",
    year = "2004"
}

@article{Dong:2009uf,
    author = "Dong, Yubing and Faessler, Amand and Gutsche, Thomas and Lyubovitskij, Valery E.",
    title = "{$J/\psi \gamma$ and $\psi(2S) \gamma$  decay modes of the $X(3872)$}",
    eprint = "0909.0380",
    archivePrefix = "arXiv",
    primaryClass = "hep-ph",
    doi = "10.1088/0954-3899/38/1/015001",
    journal = "J. Phys. G",
    volume = "38",
    pages = "015001",
    year = "2011"
}

@article{Dai:2019hrf,
    author = "Dai, Lin and Guo, Feng-Kun and Mehen, Thomas",
    title = "{Revisiting $X(3872)\to D^0 \bar{D}^0 \pi^0$ in an effective field theory for the $X$(3872)}",
    eprint = "1912.04317",
    archivePrefix = "arXiv",
    primaryClass = "hep-ph",
    doi = "10.1103/PhysRevD.101.054024",
    journal = "Phys. Rev. D",
    volume = "101",
    number = "5",
    pages = "054024",
    year = "2020"
}

@article{Mehen:2015efa,
    author = "Mehen, Thomas",
    title = "{Hadronic loops versus factorization in effective field theory calculations of $X(3872) \to \chi_{cJ}\pi^0$}",
    eprint = "1503.02719",
    archivePrefix = "arXiv",
    primaryClass = "hep-ph",
    doi = "10.1103/PhysRevD.92.034019",
    journal = "Phys. Rev. D",
    volume = "92",
    number = "3",
    pages = "034019",
    year = "2015"
}

@article{Fleming:2011xa,
    author = "Fleming, Sean and Mehen, Thomas",
    title = "{The decay of the $X(3872)$ into $\chi_{cJ}$ and the Operator Product Expansion in XEFT}",
    eprint = "1110.0265",
    archivePrefix = "arXiv",
    primaryClass = "hep-ph",
    reportNumber = "INT-PUB-11-042",
    doi = "10.1103/PhysRevD.85.014016",
    journal = "Phys. Rev. D",
    volume = "85",
    pages = "014016",
    year = "2012"
}

@article{Braaten:2010mg,
    author = "Braaten, Eric and Hammer, H. -W. and Mehen, Thomas",
    title = "{Scattering of an Ultrasoft Pion and the $X(3872)$}",
    eprint = "1005.1688",
    archivePrefix = "arXiv",
    primaryClass = "hep-ph",
    reportNumber = "INT-PUB-10-019, HISKP-TH-10-12",
    doi = "10.1103/PhysRevD.82.034018",
    journal = "Phys. Rev. D",
    volume = "82",
    pages = "034018",
    year = "2010"
}

@article{Braaten:2015tga,
    author = "Braaten, Eric",
    title = "{Galilean-invariant effective field theory for the $X(3872)$}",
    eprint = "1503.04791",
    archivePrefix = "arXiv",
    primaryClass = "hep-ph",
    doi = "10.1103/PhysRevD.91.114007",
    journal = "Phys. Rev. D",
    volume = "91",
    number = "11",
    pages = "114007",
    year = "2015"
}

@article{Jia:2023hvc,
    author = "Jia, Zhao-Sai and Zhang, Zhen-Hua and Li, Gang and Guo, Feng-Kun",
    title = "{Radiative decays of the heavy-quark-spin molecular partner of $T_{cc}^+$}",
    eprint = "2307.11047",
    archivePrefix = "arXiv",
    primaryClass = "hep-ph",
    doi = "10.1103/PhysRevD.108.094038",
    journal = "Phys. Rev. D",
    volume = "108",
    number = "9",
    pages = "094038",
    year = "2023"
}

@article{Fleming:2021wmk,
    author = "Fleming, Sean and Hodges, Reed and Mehen, Thomas",
    title = "{$T_{cc}^+$~decays: Differential spectra and two-body final states}",
    eprint = "2109.02188",
    archivePrefix = "arXiv",
    primaryClass = "hep-ph",
    doi = "10.1103/PhysRevD.104.116010",
    journal = "Phys. Rev. D",
    volume = "104",
    number = "11",
    pages = "116010",
    year = "2021"
}

@article{Amundson:1992yp,
    author = "Amundson, James F. and Boyd, C. Glenn and Jenkins, Elizabeth Ellen and Luke, Michael E. and Manohar, Aneesh V. and Rosner, Jonathan L. and Savage, Martin J. and Wise, Mark B.",
    title = "{Radiative $D^*$ decay using heavy quark and chiral symmetry}",
    eprint = "hep-ph/9209241",
    archivePrefix = "arXiv",
    reportNumber = "UCSD-PTH-92-31, CALT-68-1816, EFI-92-45, CERN-TH-6650-92",
    doi = "10.1016/0370-2693(92)91341-6",
    journal = "Phys. Lett. B",
    volume = "296",
    pages = "415--419",
    year = "1992"
}

@article{Stewart:1998ke,
    author = "Stewart, Iain W.",
    title = "{Extraction of the $D^* D \pi$ coupling from $D^*$ decays}",
    eprint = "hep-ph/9803227",
    archivePrefix = "arXiv",
    reportNumber = "CALT-68-2160",
    doi = "10.1016/S0550-3213(98)00374-5",
    journal = "Nucl. Phys. B",
    volume = "529",
    pages = "62--80",
    year = "1998"
}

@article{Guo:2019qcn,
    author = "Guo, Feng-Kun",
    title = "{Novel Method for Precisely Measuring the $X(3872)$ Mass}",
    eprint = "1902.11221",
    archivePrefix = "arXiv",
    primaryClass = "hep-ph",
    doi = "10.1103/PhysRevLett.122.202002",
    journal = "Phys. Rev. Lett.",
    volume = "122",
    number = "20",
    pages = "202002",
    year = "2019"
}

@article{Wang:2022qxe,
    author = "Wang, Yan and Wu, Qi and Li, Gang and Qin, Wen-Hua and Liu, Xiao-Hai and An, Chun-Sheng and Xie, Ju-Jun",
    title = "{Investigations of charmless decays of $X(3872)$ via intermediate meson loops}",
    eprint = "2209.12206",
    archivePrefix = "arXiv",
    primaryClass = "hep-ph",
    doi = "10.1103/PhysRevD.106.074015",
    journal = "Phys. Rev. D",
    volume = "106",
    number = "7",
    pages = "074015",
    year = "2022"
}

@article{Albaladejo:2015dsa,
    author = "Albaladejo, M. and Guo, F. -K. and Hidalgo-Duque, C. and Nieves, J. and Valderrama, M. Pavon",
    title = "{Decay widths of the spin-2 partners of the $X(3872)$}",
    eprint = "1504.00861",
    archivePrefix = "arXiv",
    primaryClass = "hep-ph",
    doi = "10.1140/epjc/s10052-015-3753-6",
    journal = "Eur. Phys. J. C",
    volume = "75",
    number = "11",
    pages = "547",
    year = "2015"
}

@article{Guo:2010ak,
    author = "Guo, Feng-Kun and Hanhart, Christoph and Li, Gang and Mei{\ss}ner, Ulf-G. and Zhao, Qiang",
    title = "{Effect of charmed meson loops on charmonium transitions}",
    eprint = "1008.3632",
    archivePrefix = "arXiv",
    primaryClass = "hep-ph",
    reportNumber = "FZJ-IKP-TH-2010-08, HISKP-TH-10-09",
    doi = "10.1103/PhysRevD.83.034013",
    journal = "Phys. Rev. D",
    volume = "83",
    pages = "034013",
    year = "2011"
}

@article{BESIII:2020nbj,
    author = "Ablikim, Medina and others",
    collaboration = "BESIII",
    title = "{Study of Open-Charm Decays and Radiative Transitions of the $X(3872)$}",
    eprint = "2001.01156",
    archivePrefix = "arXiv",
    primaryClass = "hep-ex",
    doi = "10.1103/PhysRevLett.124.242001",
    journal = "Phys. Rev. Lett.",
    volume = "124",
    number = "24",
    pages = "242001",
    year = "2020"
}

@article{Nakamura:2026xax,
    author = "Nakamura, Satoshi X.",
    title = "{Two nearby states in the $X(3872)$ region: Resolving the radiative-decay ratio tension with $\eta_{c2}$}",
    eprint = "2602.19165",
    archivePrefix = "arXiv",
    primaryClass = "hep-ph",
    doi = "10.1103/rr8b-j4nw",
    journal = "Phys. Rev. D",
    volume = "113",
    number = "11",
    pages = "114031",
    year = "2026"
}

@article{Wang:2025dur,
    author = "Wang, Xiongfei and Liu, Xiang and Gao, Yuanning",
    title = "{Colloquium: Hadron production in open-charm meson pairs at $e^+e^-$ colliders}",
    eprint = "2502.15117",
    archivePrefix = "arXiv",
    primaryClass = "hep-ex",
    doi = "10.1103/2mrp-chly",
    journal = "Rev. Mod. Phys.",
    volume = "98",
    number = "2",
    pages = "021001",
    year = "2026"
}

@article{Colangelo:2007ph,
    author = "Colangelo, P. and De Fazio, F. and Nicotri, S.",
    title = "{$X(3872) \to D \bar{D} \gamma$ decays and the structure of $X(3872)$}",
    eprint = "hep-ph/0701052",
    archivePrefix = "arXiv",
    reportNumber = "BARI-TH-07-556",
    doi = "10.1016/j.physletb.2007.05.012",
    journal = "Phys. Lett. B",
    volume = "650",
    pages = "166--171",
    year = "2007"
}

@article{DeFazio:2008xq,
    author = "De Fazio, Fulvia",
    title = "{Radiative transitions of heavy quarkonium states}",
    eprint = "0812.0716",
    archivePrefix = "arXiv",
    primaryClass = "hep-ph",
    reportNumber = "BARI-TH-08-600",
    doi = "10.1103/PhysRevD.83.099901",
    journal = "Phys. Rev. D",
    volume = "79",
    pages = "054015",
    year = "2009",
    note = "[Erratum: Phys.Rev.D 83, 099901 (2011)]"
}

@article{Colangelo:2025uhs,
    author = "Colangelo, Pietro and De Fazio, Fulvia and Roselli, Giuseppe",
    title = "{Charming case of $X(3872)$ and $\chi_{c1}(2P)$}",
    eprint = "2501.15888",
    archivePrefix = "arXiv",
    primaryClass = "hep-ph",
    reportNumber = "BARI-TH/770-25",
    doi = "10.1103/PhysRevD.111.074014",
    journal = "Phys. Rev. D",
    volume = "111",
    number = "7",
    pages = "074014",
    year = "2025"
}

@article{Voloshin:2005rt,
    author = "Voloshin, M. B.",
    title = "{$X(3872)$ diagnostics with decays to $D \bar{D} \gamma$}",
    eprint = "hep-ph/0509192",
    archivePrefix = "arXiv",
    reportNumber = "FTPI-MINN-05-42-T, UMN-TH-2415-05",
    doi = "10.1142/S0217751X06029041",
    journal = "Int. J. Mod. Phys. A",
    volume = "21",
    pages = "1239--1250",
    year = "2006"
}

@article{Kaplan:1998tg,
    author = "Kaplan, David B. and Savage, Martin J. and Wise, Mark B.",
    title = "{A New expansion for nucleon-nucleon interactions}",
    eprint = "nucl-th/9801034",
    archivePrefix = "arXiv",
    reportNumber = "DOE-ER-40561-352, INT-97-00-189, NT-UW-98-05, CALT-68-2155",
    doi = "10.1016/S0370-2693(98)00210-X",
    journal = "Phys. Lett. B",
    volume = "424",
    pages = "390--396",
    year = "1998"
}

@article{Casalbuoni:1996pg,
    author = "Casalbuoni, R. and Deandrea, A. and Di Bartolomeo, N. and Gatto, Raoul and Feruglio, F. and Nardulli, G.",
    title = "{Phenomenology of heavy meson chiral Lagrangians}",
    eprint = "hep-ph/9605342",
    archivePrefix = "arXiv",
    reportNumber = "UGVA-DPT-1996-05-928, BARI-TH-96-237",
    doi = "10.1016/S0370-1573(96)00027-0",
    journal = "Phys. Rept.",
    volume = "281",
    pages = "145--238",
    year = "1997"
}

@article{Colangelo:2003sa,
    author = "Colangelo, P. and De Fazio, F. and Pham, T. N.",
    title = "{Nonfactorizable contributions in $B$ decays to charmonium: The Case of $B^-\to K^- h_c$}",
    eprint = "hep-ph/0310084",
    archivePrefix = "arXiv",
    reportNumber = "BARI-TH-03-467",
    doi = "10.1103/PhysRevD.69.054023",
    journal = "Phys. Rev. D",
    volume = "69",
    pages = "054023",
    year = "2004"
}

@article{Guo:2013zbw,
    author = "Guo, Feng-Kun and Hanhart, Christoph and Mei{\ss}ner, Ulf-G. and Wang, Qian and Zhao, Qiang",
    title = "{Production of the $X(3872)$ in charmonia radiative decays}",
    eprint = "1306.3096",
    archivePrefix = "arXiv",
    primaryClass = "hep-ph",
    doi = "10.1016/j.physletb.2013.06.053",
    journal = "Phys. Lett. B",
    volume = "725",
    pages = "127--133",
    year = "2013"
}

@article{Guo:2019twa,
    author = "Guo, Feng-Kun and Liu, Xiao-Hai and Sakai, Shuntaro",
    title = "{Threshold cusps and triangle singularities in hadronic reactions}",
    eprint = "1912.07030",
    archivePrefix = "arXiv",
    primaryClass = "hep-ph",
    doi = "10.1016/j.ppnp.2020.103757",
    journal = "Prog. Part. Nucl. Phys.",
    volume = "112",
    pages = "103757",
    year = "2020"
}

@article{Yuan:2025pnt,
    author = "Yuan, Qi-Wei and Wu, Qi and Liu, Ming-Zhu",
    title = "{$B$-meson decays to vector charmoniumlike states and a $K$ meson: The role of final-state interactions}",
    eprint = "2504.11121",
    archivePrefix = "arXiv",
    primaryClass = "hep-ph",
    doi = "10.1103/xrjr-zxpz",
    journal = "Phys. Rev. D",
    volume = "111",
    number = "11",
    pages = "114035",
    year = "2025"
}

\end{document}